\documentclass[12pt]{article}
\usepackage{amssymb}
\usepackage{booktabs}
\usepackage{multirow}
\usepackage{soul}
\usepackage{upgreek}
\usepackage{microtype}
\usepackage{graphicx}
\usepackage{tabularx}
\usepackage{hyperref}

\title{Angular reconstruction of high energy air showers using the radio signal spectrum}
\author{S Nonis$^2$, A Leisos$^1$,
A Tsirigotis$^1$, G Bourlis$^1$, K Papageorgiou$^2$,\\ I Gkialas$^2$, I Manthos$^3$  and Spyros Tzamarias$^3$}
\date{%
$^1$ Physics Laboratory, School of Science and Technology, Hellenic Open University, Patras 26222, Greece \\
$^2$ School of Engineering, Department of Financial and Management Engineering, University of the Aegean, Chios 82100, Greece\\
$^3$ Department of Physics, Aristotle University of Thessaloniki, Thessaloniki 54124, Greece\\
}
\begin{document}
\maketitle

\begin{abstract}
The Hellenic Open University extensive air shower array (also known as Astroneu array) is a small scale hybrid detection system operating in an area with high levels of electromagnetic noise from anthropogenic activity. In the present study we report the latest results of the data analysis concerning the estimation of the shower direction using the spectrum of the RF system. In a recent layout of the array, 4 RF antennas were operating receiving a common trigger from an autonomous detection station of 3 particle detectors. The directions estimated with the RF system are in very good agreement with the corresponding estimations using the particle detectors demonstrating that a single antenna has the potential for reconstructing the shower axis angular direction.
\end{abstract}

%\submitto{\PS}

\section{Introduction}\label{sect1}
\paragraph{}The detection of high energetic cosmic particles (above $10^{14}$ eV) is traditionally made by well-established techniques which use ground particle detectors  or detectors recording Cherenkov and fluorescent light emission. Another method of  detection relies on  the measurement of the electromagnetic radiation emitted by high energy showers in the radio frequency (RF) regime. The first RF signal detection from air showers was made during the 1960's \cite{1}, but  the lack of fast digital electronics has soon sidelined this method. Since 2002 the RF detection has been regenerated from various experiments (such as LOPES \cite{2}, CODALEMA \cite{3}, LOFAR \cite{4}, AERA \cite{5} and Tunka-Rex \cite{6}) that measure the electric field strength at frequency range [30-80] MHz (or up to the band [110-200] MHz). Nowadays, the RF detection allows the measurements of properties of   primary cosmic particles, such as their arrival direction, energy and chemical composition with accuracy similar to those obtained by particle detectors or fluorescence telescopes. Among the advantages of the RF method is the small dependence on the atmospheric conditions (weather, light and transparency), as well as the low cost of the antennas (and their electronics) compared to large scintillator detectors.

The radio emission in air showers is attributed to two different physical processes. The more dominant  is of geomagnetic origin,  producing a time-varying transverse (with respect to the shower axis) current, due to the (opposite) deflection of shower's electrons and positrons by the Earth's magnetic field \cite{7}. The RF signal generated by this mechanism is linearly polarized in the direction of the Lorentz force ($\vec{v}\times\vec{B}$). A secondary contribution to the radio signal comes from the excess of electrons at the front of the shower (Askaryan effect) \cite{8}, which generates a time depended current collateral to the shower axis direction. The resulting signal is radially polarized, pointing towards the shower axis. The RF electric field measured at the ground is the sum of both contributions. 

The Hellenic Open University (HOU) extensive air shower array (Astroneu \cite{9}) is a hybrid small scale array, operating in urban environment,  on the outskirts of the city of Patras in Greece. During the first pilot phase (2014-2017) the array consisted of 3 stations each comprising 3 large scintillator detectors and 1 RF antenna. Since 2017 (second phase) 4 RF antennas were deployed at station A, receiving a common trigger from the 3 scintillator detectors.
In former studies, we demonstrated the performance of the Astroneu array with emphasis on the detection and reconstruction of EAS using the charged particle detectors \cite{9} indicating also the capability of detecting RF signals from showers  by imposing the appropriate selection criteria to the RF signals \cite{10}. We have also studied the timing and the amplitude strength of the RF signals by comparing the antenna data with the particle detector data, as well as with the simulation predictions \cite{11}. Furthermore, we have presented the first studies on a complete Voltage Response Model (VRM) for the RF system using the antenna's Vector Effective Length (VEL) \cite{12} and the measured electric field (actually the RF spectra) at the antenna's position in order to estimate the primary particle arrival direction.

In this work we extend this study using a specific geometrical layout and we compare our RF measurements to independent measurements obtained by the particle detectors as well as to the simulation predictions.
%The paper is organized as follows;
In Section 2 we outline the components and  performance of the Astroneu array, while in Section 3 the simulation framework is presented. In Section 4 we describe the method for  reconstructing the direction of the shower axis using the RF spectrum and evaluate its performance using Monte Carlo (MC) simulation. In Section 5 we apply the method to  shower data collected by the Astroneu array and we compare the estimations of the angular shower axis direction using the RF spectrum with the independent measurements of the particle detectors. Finally, in Section 6 the conclusions are drawn.

\section{The Astroneu array}{\label{sect2}}

\begin{figure}[h!]
\centering
\includegraphics[width=\textwidth]{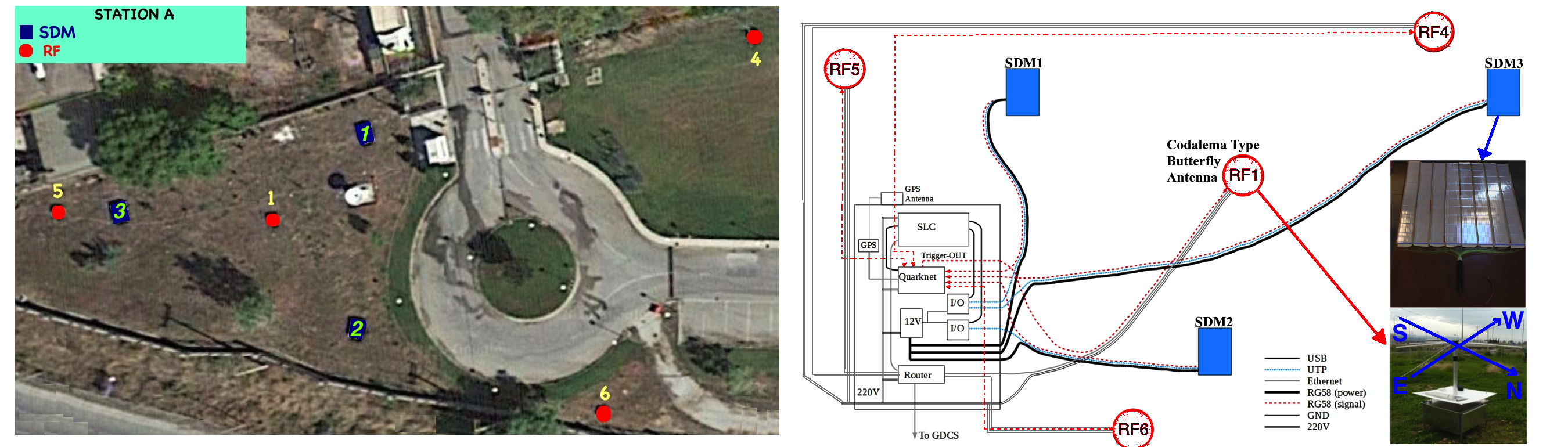} 
\caption{(a) ({\it Left}) The layout of Astronue's station-A. The 3 SDMs are marked with blue squares, while the 4 RF antennas with red circles. (b) ({\it Right}) Schematic description of the connections in  station-A. }
\label{fig1}
\end{figure}

The Astroneu array is a small scale hybrid detection system comprising 9 scintillator detectors and 6 RF antennas. The detector components are arranged in three independent autonomous stations (A, B and C) separated  by few hundred meters, while the inter-station distances between the particle detectors is about 27 meters. 
Each station includes three Scintillator Detector Modules (SDM) \cite{9} forming a triangle  with one RF Antenna (RFA) \cite{13} in the middle.  It is also equipped with trigger, digitization and Data Acquisition (DAQ) electronics along with slow control and monitor electronics and a GPS-based timing system. 
Since 2017 station A comprises 3 more antennas (4 RFAs in total)  covering an area roughly 120\,m in diameter.
The data used in this analysis were collected by station A of which the geometrical layout is depicted in Figure \ref{fig1} (left). 

The SDM consists of $160$ scintillating tiles covering an area of approximately $1\,m^{2}$. The light which is generated during the interaction of shower particles with the scintillation material is guided to a single Photomultiplier tube (PMT) using $96$ embedded wavelength shifting fibers (WLS). The RFA is a "Butterfly" bowtie antenna \cite{13}  designed and constructed by the CODALEMA collaboration \cite{3}.  The antenna comprises  two orthogonal electrically insulated dipoles oriented in the East-West (EW) and North-South (NS) directions. The dipole signals are led directly into the input of a Low Noise Amplifier (LNA) \cite{13} (with internal impedance which matches the antenna characteristics at frequency range [20-80] MHz) at the center of the antenna. Each antenna is equipped with trigger and digitization electronics along with a GPS based timing system. Upon a trigger, provided by the particle detectors, the waveforms of the two dipoles are sampled at a rate of 1 GHz. 
A schematic representation of the Astroneu's station-A connections is depicted in Figure \ref{fig1} (right). 

The PMT signals of the three SDM  are acquired by a Quarknet electronics board \cite{14} that digitizes ( $1.25 \,ns$ resolution) the crossing of the waveforms with a predefined voltage threshold. The occurrence  of the first crossing is used for timestamping while the period of time that the pulse remains above the threshold   (Time over Threshold - ToT) is used for  pulse size estimation \cite{calib}. Absolute timing is provided by a GPS receiver.
%A Quarknet electronics board \cite{15} is used for reading the three SDM PMTs signals of the station by measuring up to $1.25$ $ns$ accuracy the passing time of the waveforms with a predetermined voltage minimum (threshold). The time of first pass determines the timing of the pulse, while the time that the pulse exceeds the voltage minimum (Time over Threshold - ToT) is used for the pulse size estimation.
The necessary I/O and network devices, the Quarknet board as well as the Station Local Computer (SLC)  are placed inside a metallic Central Electronics Box (CEB). A station trigger is formed when all three SDMs of the station have signals exceeding $9.7 \,mV$ in a time window of $240\,ns$. This trigger signal is fed into the RFA external trigger input which starts the recording of the EW and NS electric-field waveforms. The signals are digitized by an ADC (1\,GS/s over 2560 points for 2.56\,s record) and timestamped using GPS. The experimental data from both the SDMs and the RFAs of the station are transferred to the Global Data and Control Server (GDCS), where the event building is carried out offline using  the GPS time-tags. 

Each autonomous station of the Astroneu array can potentially reconstruct extensive air showers of energy more than $10 \,TeV$ with a typical resolution of $3.5$ degrees at a rate of $17 \,h^{-1}$. The efficiency of the Astroneu array in detecting and reconstructing EAS using the data from the charged particle detectors of one station (single station operation) or by combining the data information from two stations (multiple station operation) is reported in \cite{9}. The RF component of the EAS has been studied using noise filters, timing and signal polarization \cite{10}. Further studies including the correlation of the RF signals with the particle detector data as well as the comparison of the electric field measurements with the MC prediction have also been reported \cite{11}.

\section{Simulation framework}{\label{sect3}
The simulation procedure can be divided into three main parts. In the first simulation step we  produced high energy shower events (corresponding to $348000$ hours of experimental time) with the CORSIKA simulation package \cite{15}. The energy varied between $10^{15} eV$ and $10^{18} eV$ with primary relative abundances and spectral index according to the latest measurements \cite{16}. The QGSJET-II-04 \cite{17} package has been utilized for the hadronic interactions and the EGS4 Code system \cite{18} for the electromagnetic interactions. The simulations sample was selected to cover a wide range of arrival directions ($0^{o}\leq\theta\leq90^{o}, 0\leq\phi\leq360^{o}$) and showers cores inside a large enough circular area of radius $R=418m$ around the center of station-A. During the second part of the simulation process, the Hellenic Open University Reconstruction and Simulation (HOURS) package \cite{19} was applied to simulate the response of the SDM to shower particles (scintillating material, optical fibers, photomultipliers and cable effects) and the functionality of the trigger and data acquisition system.
% recording for the simulated showers the ToT in each SDM. 

The final stage in the simulation process is the RF signal development.
For the simulation of the RF signals the Simulation of Electric Field from Air Showers (SELFAS) package \cite{20} was used, which calculates the electric field of the RF  generated signal during the shower  development in the atmosphere. SELFAS takes into account the dominant  contributions by transverse current variation and by charge excess variation. In order to calculate the voltage induced on the
EW/NS terminals of the antenna when an electromagnetic wave coming from a given direction ($\theta, \phi$) impinges on it, the convolution of the wave field with the antenna's vector effective length (VEL) is commonly used. The VEL of an antenna  can be expressed in terms of the gain and structural features of the antenna (i.e., the antenna radiation resistance and reactance) as well as on the Low Noise Amplifier's characteristics (e.g., input resistance and reactance) as described in \cite{21}. The computation of the VEL is acquired with the Numerical Electromagnetic Code (NEC) and the 4NEC2 software \cite{22}. It should be emphasized that during VEL's calculations the far field approximation  is taken into account. This implies that we measure the response of the antennas by considering the plane wavefront approximation. 

After  convolution, the RF signal is distorted with noise with an average rms of $7.61\,mV$ for antenna 1, $4.8\,mV$ for antenna 4, $12.61\,mV$ for antenna 5 and $6.39\,mV$ for antenna 6, as calculated from a large number  of background events for the 4 antennas of station-A. Then the analysis follows the standard reconstruction algorithms which are applied either to simulated events or experimental data.
%is added the simulated RF signal we add RF noise as calculated for a large number ($10.000$) of noisy events of the 4 antennas of station-A. The average noise estimated to be $7.61mV$ for antenna 1, $4.8mV$ for antenna 4, $12.61mV$ for antenna 5 and $6.39mV$ for antenna 6. The following analysis steps were applied to both the experimental data and the simulated events.

\section{Estimation of the shower axis direction using the RF spectrum}{\label{sect4}}
\paragraph{} The antenna's VEL, for EW and NS directions is a
vector quantity and can be formulated with a two-component vector along the directions of the unit vectors $\bf{e}_{\theta}$ and $\bf{e}_{\phi}$, as indicated in Figure \ref{fig2}.
\begin{figure}[h!]
\centering
\includegraphics[scale=0.4]{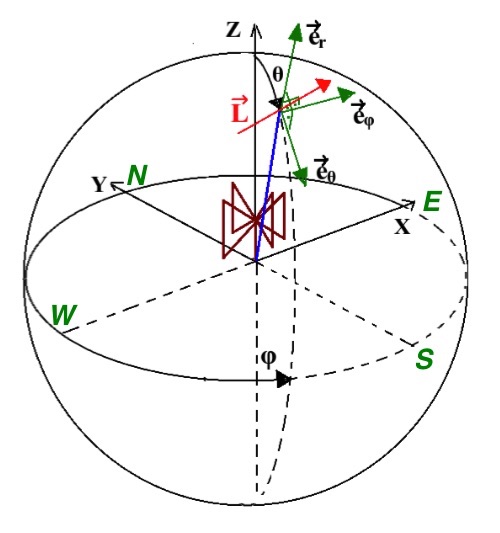} 
\caption{A spherical coordinate system used for the antenna's VEL calculations. The antenna is located at the origin of the system. The reference point for measuring the zenith angle $\theta$ is the top one, while the azimuth angle $\phi$ is counted from the x-axis (along EW direction) of system. The electromagnetic wave arrives from the direction ($\theta,\phi$). The VEL $\bf{L}$ for the wave's arrival direction is drawn with red color.} 
\label{fig2}
\end{figure}	 
Assuming an incident electromagnetic field from the (zenith=$\theta$, azimuth=$\phi$) direction, the induced voltage across the arms of the antenna (EW or NS dipole) is obtained by  convoluting the electric field and the VEL of the antenna \cite{23}:
%The voltage developed across the arms of the antenna (including that on the LNA) created by the impact of an electromagnetic field of known direction (zenith=$\theta$, azimuth=$\phi$) (in a single polarization) is obtained by the convolution of the field and the VEL of the antenna.
%For the EW/NS antenna's poles the induced voltage can be summarized \cite{23} in the following expression
\begin{equation}
V_{EW/NS}(\theta, \phi, t) = \mathbf{L}_{EW/NS}(\theta, \phi, t) \ast \mathbf{E}_{EW/NS}(\theta, \phi, t)
\label{eq1}
\end{equation}
In the frequency domain the antenna response (Figure  \ref{fig3}) can be expressed as the dot product of their Fourier transforms (according to convolution theorem)
\begin{equation}
V_{EW/NS}(\theta, \phi, f ) = \mathbf{L}_{EW/NS}(\theta, \phi, f ) \cdot \mathbf{E}_{EW/NS}(\theta, \phi, f )
\label{eq2}
\end{equation}
\begin{figure}[h!]
\centering
\includegraphics[width=\textwidth]{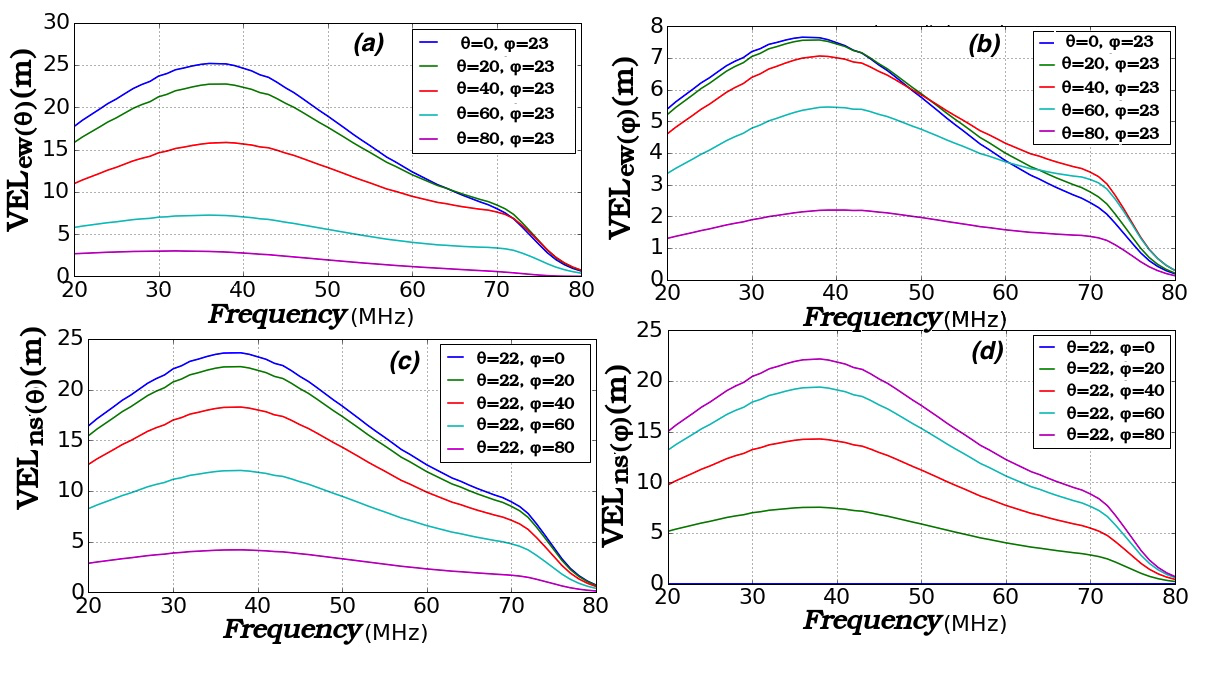} 
\caption{The frequency dependence of antenna's VEL components. {\it (a), (b)} The $\theta$-component and $\phi$-component of VEL respectively for different zenith angles. {\it (c), (d)} The $\theta$-component and $\phi$-component of VEL respectively for different azimuth angles.} 
\label{fig3}
\end{figure}	 
while the RF power spectrum  $P_{EW/NS}(\theta,\phi,f)$ can be calculated using the formula: 
\begin{equation}
P_{EW/NS}(\theta,\phi,f)=20\log{\left(\frac{\left|V_{EW/NS}(\theta,\phi,f)\right|}{\sqrt{Z_{sys}\cdot 10^{-3}}}\right)}
\label{eq4}
\end{equation}
where $Z_{sys}$ is the system (antenna+LNA) impedance.

%The frequency dependence of the antenna's VEL is presented in Figure \ref{fig3}.
The dependence of the VEL (and consequently of the RF power spectrum) on the azimuth and zenith angle of the primary particle's direction shown in Figure \ref{fig3} suggests that  
the arrival direction can be estimated by comparing the spectrum $P_{EW/NS}(f)$ of the observed event with the spectrum $P_{EW/NS}(\theta,\phi,f)$ of the antenna response model. In this analysis the $P_{EW/NS}(\theta,\phi,f)$ response spectrum was computed for a sufficiently large number of $(\theta,\phi)$ pairs (i.e. for $0^{0}\leq\theta\leq90^{0}$, $0^{0}\leq\phi\leq 360^{0}$  with a step of one degree for $\theta$, and a half degree for $\phi$). 
%The RF power spectrum $P_{ew/ns}^{spec}(f)$ (in dB) as well as the model's power spectrum $P_{ew/ns}(\theta,\phi,f)$ can be calculated using the formulas 
%\begin{equation}
%P_{ew/ns}^{spec}(f)=20\log{\left(\frac{\left|V_{ew/ns}^{spectrum}(f)\right|}{\sqrt{Z_{sys}\cdot 10^{-3}}}\right)}
%\label{eq3}
%\end{equation}
%\begin{equation}
%P_{ew/ns}(\theta,\phi,f)=20\log{\left(\frac{\left|V_{ew/ns}(\theta,\phi,f)\right|}{\sqrt{Z_{sys}\cdot 10^{-3}}}\right)}
%\label{eq4}
%\end{equation}
%where $Z_{sys}$ is the system (antenna+LNA) impedance.
Then the shower direction $(\hat{\theta},\hat{\phi})$ was estimated by minimizing the quantity:
\begin{equation}
\chi^2(\theta,\phi)=\sum_{[30-80]MHz}{\left(P_{EW/NS}^{model}(\theta,\phi,f)-P_{EW/NS}(f)\right)^2}
\label{eq5}
\end{equation}
where the summation is implemented over the frequency values [30-80]\,MHz (a step of 2 MHz was used) and for both poles of the antenna, while the model power spectra for  any $(\theta,\phi)$ values  were obtained using linear interpolation.

\begin{figure}[h!]
\centering
\includegraphics[scale=2.44]{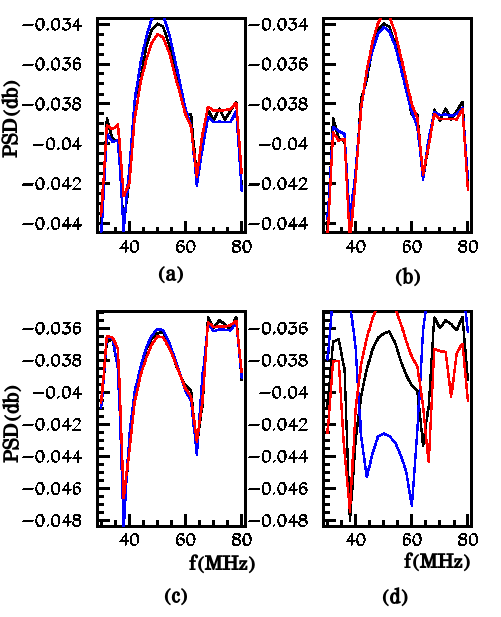}
\caption{{\it (a)} The EW spectrum of an event recorded by an antenna (black line).The blue line corresponds to the model spectrum with $(\theta,\phi)=(\hat{\theta}+5^{o},\hat{\phi})$ while the red one to  $(\theta,\phi)=(\hat{\theta}-5^{o},\hat{\phi})$ . {\it (b)} The black line represents the same spectrum as in case {\it (a)} while the blue line corresponds to the model spectrum with $(\theta,\phi)=(\hat{\theta},\hat{\phi}+15^{o})$ and the red one to $(\theta,\phi)=(\hat{\theta},\hat{\phi}-15^{o})$. {\it (c)} and {\it (d)} similar to (a) and (b) respectively for the NS direction.}   
\label{fig4}
\end{figure}
It should be noted that the spectra used in equation \ref{eq5} are normalized to unity (with respect to the frequency) in order to compensate the attenuation of the electric field with the distance from the shower axis and the energy of the primary particle. Consequently, in this analysis the power spectra are assumed to be independent from any other physical characteristic of the shower except the direction of the primary particle. 
For example, Figure \ref{fig4} presents the power spectrum of a recorded event in comparison with the model spectra. The shown model spectra correspond to $(\theta,\phi)$ pairs near the estimated values $(\hat{\theta},\hat{\phi})$ i.e. $(\theta,\phi)=(\hat{\theta}\pm5^{o},\hat{\phi})$ and  $(\theta,\phi)=(\hat{\theta},\hat{\phi}\pm15^{o})$ for both directions EW and NS.

%Applying the method to a real event the reconstructed shower direction estimated in zenith and azimuth angle $\hat{\theta}$ and $\hat{\phi}$ respectively. Figure \ref{fig4} focuses on the sensitivity of the spectrum to changes in zenith or azimuth angle. In figure \ref{fig4}(a) is represented the spectrum of the EW response model of the antenna 5 (black line) for $\hat{\theta}$, $\hat{\phi}$ while the blue and the red line correspond to the spectrum with the same azimuth angle $\hat{\phi}$ and zenith angle $\hat{\theta}+5^{o}$ and $\hat{\theta}-5^{o}$ respectively. In figure \ref{fig4}(b) the blue and the red line correspond to the spectrum with the same zenith angle $\hat{\theta}$ and azimuth angle $\hat{\phi}+15^{o}$ and $\hat{\phi}-15^{o}$ respectively. The same results for the NS spectrum of the model are depicted in figure \ref{fig4}(c,d).     

\begin{figure}[h!]
\centering
\includegraphics[scale=.5]{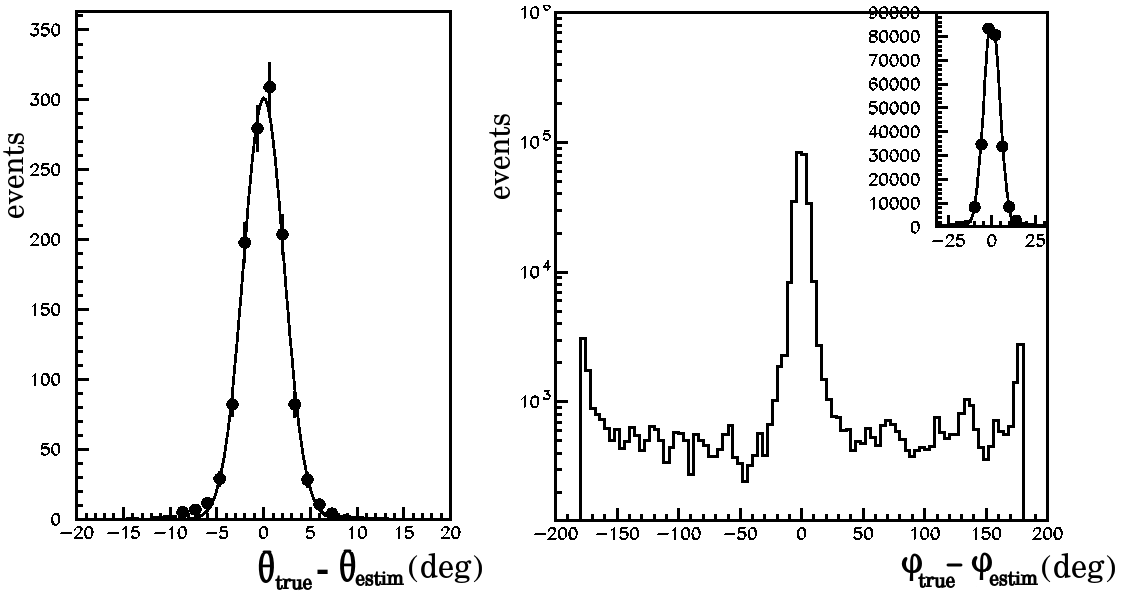} 
\caption{{\it (Left)} The distribution of $\theta_{true}-\hat{\theta}$ between the true and the estimated (using the RF spectrum) zenith angle. The distribution is fitted with Gaussian function of sigma equal to $2.2^{0}$. {\it (Right)} The distribution of $\phi_{true}-\hat{\phi}$ between the true and the estimated azimuth angle. The distribution consists of a central region which is fitted with Gaussian function of sigma equal to $5.4^{0}$ and long tails. In the inset plot linear scale is used. }
\label{fig5_a}
\end{figure}
In order to examine the sensitivity of the power spectra on the estimation of the shower axis direction, the minimization procedure (eq \ref{eq5}) was applied to a large set of MC samples (see Section 2). Figure \ref{fig5_a}(left) presents the distribution $\theta_{true}-\hat{\theta}$ between the true zenith angle and the  estimated zenith angle using the RF spectrum. The distribution is well fitted  with a Gaussian function of sigma equal to $2.2^{o}$.  In contrast to zenith angle the distribution of the difference $\phi_{true}-\hat{\phi}$ exhibits long tails while the central region around zero  fits quite well with Gaussian function of sigma equal to $5.4^{o}$ (Figure \ref{fig5_a}(right)). These  tails are highlighted in the logarithmic scale diagram while they are not visible in the linear scale diagram (inset plot of Figure \ref{fig5_a}(right)).

\begin{figure}[h!]
\centering
\includegraphics[scale=.5]{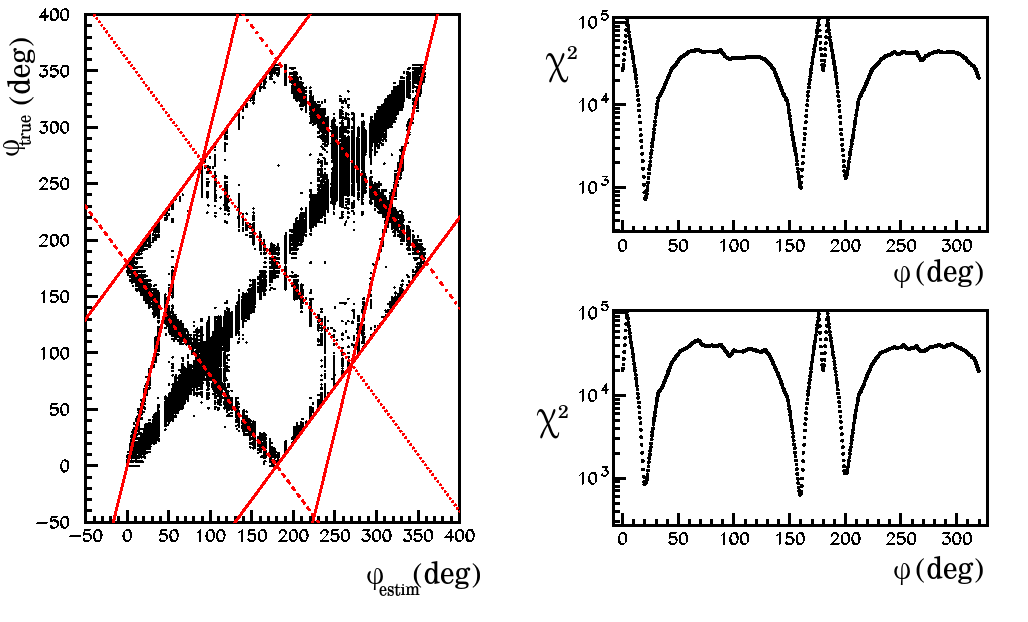} 
\caption{ {\it (Left)} The scatter plot between true and estimated azimuth angle. See text for explanation.{\it (Right)} The $\chi^2$ as function of the azimuth angle for the same event and two different antennas. In one of them (top) the minimum appears at $\hat{\phi_{1}}=20^{o}$ while in the other (bottom) at $\hat{\phi_{2}}=180-\hat{\phi_{1}}=160^{o}$. }
\label{fig5_b}
\end{figure}
The tails of Figure \ref{fig5_a}(right) are a consequence of  the symmetry of the power spectrum with respect to the azimuth angle. This is obvious if we examine the scatter plot between the estimated and  true azimuth angle presented in  Figure \ref{fig5_b}(left). Apart from the area near the  diagonal  where most of the points are located, other areas are emerged  exhibiting a symmetry pattern (highlighted with red lines). This effect is attributed to symmetries in the VEL of the antenna that previous studies have also  reported \cite{21}. For a given zenith angle $\theta$ the VEL for azimuth angles $\phi$, $(180^{o}\pm\phi)$ and $(360^{o}-\phi)$ have the same value. In addition, we have to mention some extra symmetries which concern the shape of the VEL. For example the VEL for a given pair $(\theta,\phi)$ and the VEL for $(\theta,3\phi)$ differ by a scale factor (i.e $VEL(\theta,\phi)=aVEL(\theta,3\phi)$). 
Due to these symmetries, multiple minima appear on the $\chi^{2}$ function of equation \ref{eq5} as it is clearly seen in Figure  \ref{fig5_b}(right). In this Figure, the  $\chi^2$ as a function of the azimuth angle is shown for two different antennas that detected the same event. Even if the antennas responded to the same shower, in one of them the minimum appears at $\hat{\phi_{1}}=20^{o}$ while in the other at $\hat{\phi_{2}}=180-\hat{\phi_{1}}=160^{o}$.

%As we compare the shape of the spectrum in the process of $\chi^2$ minimization, all the symmetries of the VEL are highlighted, and the azimuth angle estimation is affected. In figure \ref{fig5}(d) is shown the $\chi^2$ as a function of the azimuth angle for the same event in two different antennas. In one antenna the minimum appears in $\phi=20^{o}$ while in the other in $180-\phi=160^{o}$. 

%The high energy simulated events have been reconstructed, regarding the arrival direction, according the method decribed above. The distribution $\theta_{true}-\theta_{RF}$ between the true zenith angle and the corresponding estimated using the RF spectrum is fitted with Gaussian function of sigma equal to $2.2^{0}$ (figure \ref{fig5}(a)). In contrast to zenith angle the distribution of the differences $\phi_{true}-\phi_{estim}$ in azimuth includes a central region around zero which fits quite well with Gaussian function of sigma equal to $5.4^{0}$, but also some tails appear around the central region (figure \ref{fig5}(b)). These  tails are highlighted in the logarithmic scale diagram while they are not visible in a simple diagram (inset plot figure \ref{fig5}(b)). 

\begin{figure}[h!]
\centering
\includegraphics[scale=1.6]{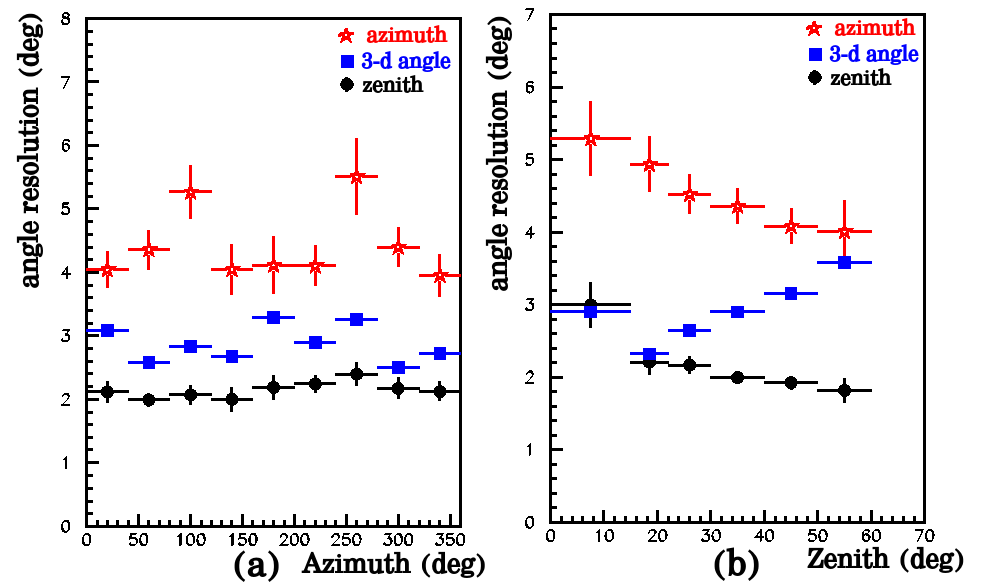} 
\caption{{\it (Left)} The resolution in estimating the zenith (black circles), azimuth (red stars) and 3-d angle (blue rectangular boxes) as a function of the azimuth angle. The 3-d angle is the angle between the true shower direction and the direction estimated from the RF signal. {\it (Right)} The resolution in estimating the zenith (black circles), azimuth (red stars) and 3-d angle (blue rectangular boxes) as a function of the zenith angle.}
\label{fig6}
\end{figure}	 

%Figure \ref{fig5}(c) displays the true azimuth angle (y-axis) and the estimated using the RF spectrum (x-axis). Apart from the central area where most of the estimated azimuth angles are located, some lines (indicated with red color) are formed in which several estimated angles are placed. This can be interpreted due to symmetries in the VEL of the antenna to which previous studies have been reported \cite{21}. For a given zenith angle $\theta$ the VEL for azimuth angles $\phi$, $(180^{o}\pm\phi)$ and $(360^{o}-\phi)$ has same value. In addition, we have to mention some extra symmetries which concern the shape of the VEL. For example the VEL for a given pair $(\theta,\phi)$ and the VEL for $(\theta,3\phi)$ differ by a scale factor (i.e $VEL(\theta,\phi)=aVEL(\theta,3\phi)$). As we compare the shape of the spectrum in the process of $\chi^2$ minimization, all the symmetries of the VEL are highlighted, and the azimuth angle estimation is affected. In figure \ref{fig5}(d) is shown the $\chi^2$ as a function of the azimuth angle for the same event in two different antennas. In one antenna the minimum appears in $\phi=20^{o}$ while in the other in $180-\phi=160^{o}$. 

In Figure \ref{fig6} is depicted the resolution in estimating the zenith and the azimuth angle of the particle initiating the shower, as well as the median of the angle between the true shower direction and the direction calculated from the RF signal (3-d angle), as a function of the zenith (a) and azimuth angle (b). The resolution in $\theta$ is evaluated as the sigma of the gaussian that fits the distribution $\theta_{true}-\hat{\theta}$, while in $\phi$ the sigma of the gaussian that fits the central part of the distribution $\phi_{true}-\hat{\phi}$.
% It is also shown the angle  between the true shower direction and the direction calculated from the RF signal (3-d angle)
%and the 3-d angle with the azimuth and zenith angles respectively. The 3-d angle is the angle between the true shower direction and the direction calculated from the RF signal. The resolution represents the square root of the variance of the antenna measurements for the same shower event. The zenith and azimuth resolution increase in the area around $\phi=90^{o}$ and $\phi=270^{o}$ while the resolution in $\theta$ and $\phi$ decrease with the zenith angle. The estimated error in both cases increases because the spectrum shape of the antenna model exhibits smaller variations.
%as indicated from
%We note that in both cases the antenna's spectrum model is denser (the %spectrum for nearby angles is similar) around these areas and the %estimation of the angles contributes to a larger errors.
\section{Application to shower data}{\label{sect6}}
\subsection{Data Selection}
\begin{figure}[h!]
\centering
\includegraphics[scale=0.5]{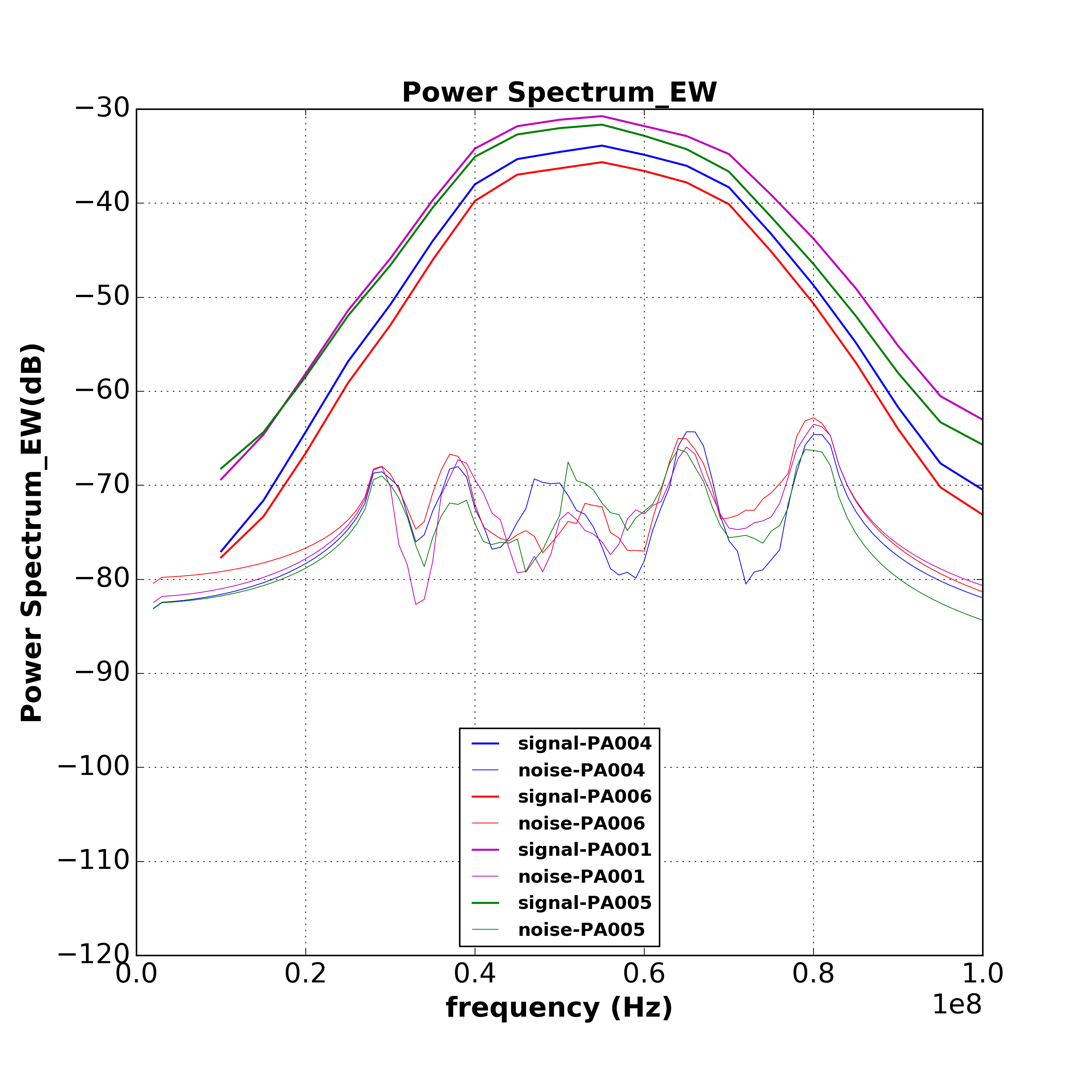} 
\caption{The EW power spectrum in the frequency range [30-80] MHz for the 4 antenna signals  for the same shower event (top lines) and the corresponding noise (bottom lines). }
%(b) ({\it Right}) The events distributions with $p\gtreqless 0.85$ (up) and rising time $\gtreqless 28ns$ (down).}
\label{fig7}
\end{figure}	
The data sample used in this analysis consisted of $470$ events collected by station-A. In order to compare the response of many antennas to the same shower, a 4-fold coincidence criterion was applied, requiring all 4 antennas to register simultaneously a signal. This criterion reduced the initial sample to $385$ events. 
For each event, all the RF signals (of both polarizations) were filtered (as described in \cite{12}) in order to keep frequencies in the range 30-80 MHz, the most appropriate frequency range for the measurement. In frequencies smaller than 20 MHz the ionospheric noise increases significantly reaching values approximately $(0.6-1.2) μV·m^{-1}·MHz^{-1}$ which are comparable to the expected electric fields values from showers. Additional at frequencies below 30 MHz the amplitude of galactic noise is about $(1-2)μV·m^{-1}·MHz^{-1}$, while at frequencies, over $80 MHz$, strong signals from the radio FM band are expected near the urban web. In Figure \ref{fig7} is shown the EW power spectrum in the frequency range [30-80] MHz for an event detected by all four antennas. 

\begin{figure}[h!]
\centering
\includegraphics[width=\textwidth]{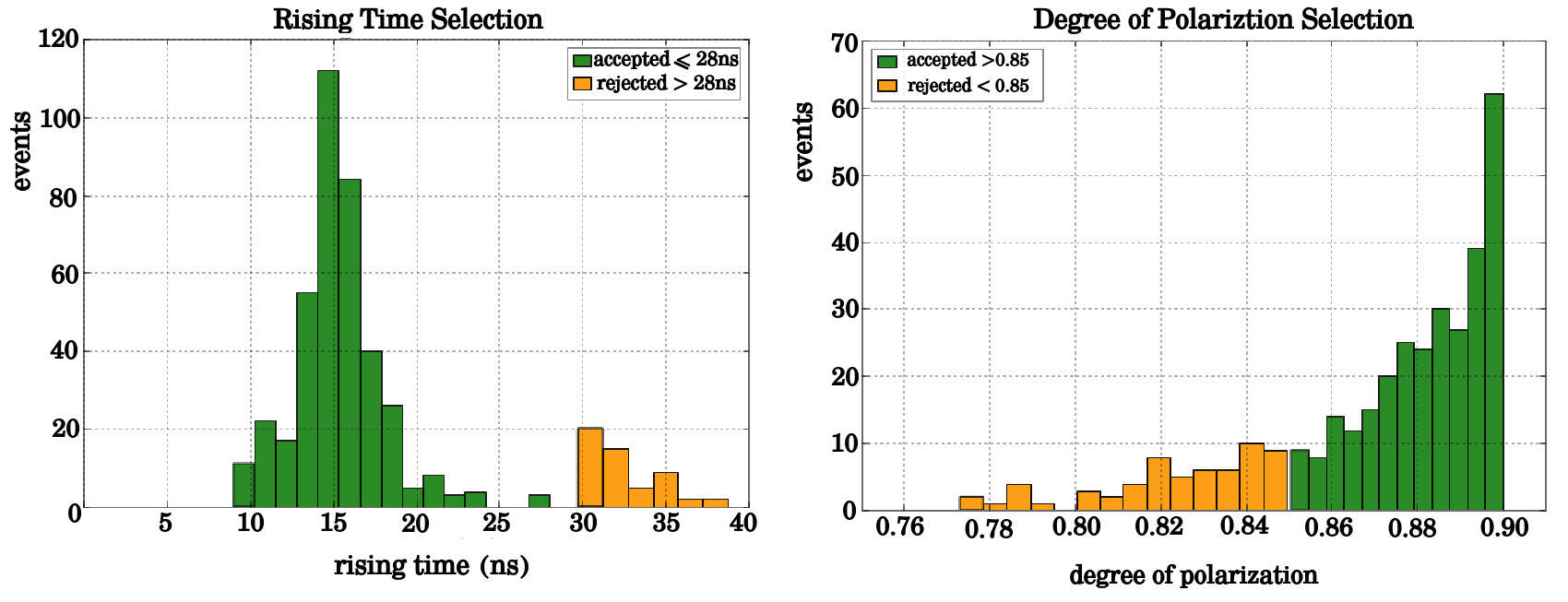} 
\caption{The distributions of the rising time (left) and the degree of polarization (right) of the sample. A cut value of 28 ns and 0.85 was applied respectively. Selected events are depicted with green color, while rejected events with orange.}
\label{fig8}
\end{figure}
For the remaining sample of $385$ events, we applied the criteria described in \cite{11} for the RF pulse period,  pulse rising time and  signal polarization for both EW and NS directions. In a brief description we can recall that the RF pulse signal from a cosmic event is expected to be around $30ns$ while the rising time of the normalized cumulative function $C(k)$ is less than $28ns$. The normalized cumulative function is defined 
\begin{equation}
C(k)=\frac{\sum\limits_{i-128}^{i+128+k}{E_j^2}}{\sum\limits_{j=i-128}^{i+128}{E_j^2}}
\label{eq6}
\end{equation}
where $i$ is the buffer position of the pulse maximum value, while $E_{j}$ is the electric field value at the buffer position $j$. The rising time of $C(k)$ is defined as the amount of time needed to increase its value from 10\% of its maximum to 80\% of its maximum. Finally the signal polarization for a cosmic shower event is predicted to be linear. We retained only those events whose the RF pulse had a degree of polarization ($p$)\footnote{$p=\frac{\sqrt{Q^2+U^2+V^2}}{I}$, where $Q,U,V,I$ the Stokes parameters of the electromagnetic signal.} greater than $0.85$ (the value $p=1$ for perfectly linear polarized wave). After all these selection criteria our final sample consisted of $274$ events for the next steps of the analysis. In Figure \ref{fig8}  is presented the distributions of the rising time and degree of polarisation before the application of the above criteria.
% events which were rejected because of the high rising time and the low degree of polarization.

\subsection{Correlation with the particle detector data}{\label{sect6}}

The reconstruction of the shower axis direction (zenith and azimuth angle) from the particle detectors was performed using the triangulation method \cite{9}. Furthermore, as described in the previous sections, we calculated the shower axis direction using the power spectra of the 4 antennas. In order to increase the resolution of the RF system, the data from the 4 antennas were combined i.e. that $\chi^{2}$ function that was minimized was:
\begin{equation}
\chi^2(\theta,\phi)={\sum_{i=1,4}{\sum_{[30-80]MHz}{\left(P_{EW/NS}^{model}(\theta,\phi,f)-P_{EW/NS}^{i}(f)\right)^2}}}
\label{eq55}
\end{equation}

%  Summarizing we have reconstructed the shower axis direction from SDM's timing and the RF spectrum, both in the events and in the MC simulations.

\begin{figure}[h!]
\centering
\includegraphics[width=\textwidth]{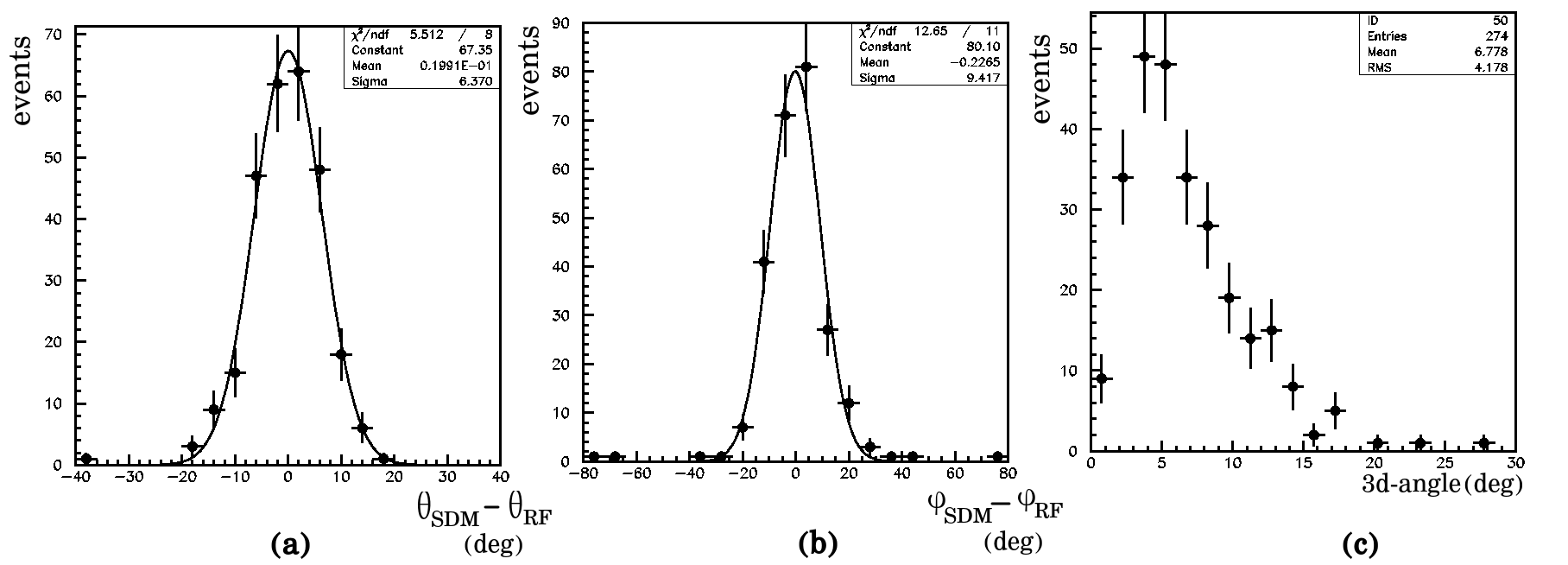} 
\caption{{\it (a)} The distribution of $\theta_{SDM}-\theta_{RF}$ between the zenith angle estimated using the SDM timing and the corresponding angle estimated using the RF spectrum. The distribution is fitted with Gaussian function of sigma equal to $6.4^{0}$. {\it (b)} The distribution of $\phi_{SDM}-\phi_{RF}$ between the azimuth angle estimated using the particle detector data and the corresponding angle estimated using the RF spectrum. The distribution is fitted with Gaussian function of sigma equal to $9.4^{0}$. {\it (c)} The distribution of the angle between the shower direction reconstructed using the SDM data and using the RF spectrum (3d-angle).} 
\label{fig9} 
\end{figure}	 
The difference between the direction estimated using the RF spectrum and the SDM timing is shown in Figure \ref{fig9}. The difference of the zenith angles (a) and azimuth angles (b)  are well described by Gaussian functions centered near zero. The corresponding sigmas of the Gaussian function are $\sigma_{\Delta\theta} \simeq 6.3^{o}$ and $\sigma_{\Delta\phi} \simeq 9.4^{o}$. In Figure \ref{fig9}(c) the distribution of the angle between the shower direction reconstructed using the SDM data and using the RF spectrum (3-d angle) is also presented. The median of this distribution is $5.8^{o}$. 

%The distributions of the difference on the zenith angle and on the azimuth angle estimated using the RF spectrum and the SDM timing are shown in figure \ref{fig9} on (a) and (b) respectively. For the angular reconstruction with the RF signal all the 4 antenna's spectrum were used in the $\chi^2$ minimization (equation \ref{eq5}). Both distributions are well described by Gaussian functions centered near zero. The corresponding sigmas of the Gaussian function are $\sigma_{\Delta\theta} \simeq 6.3^{o}$ and $\sigma_{\Delta\phi} \simeq 9.4^{o}$. In figure \ref{fig9}(c) the distribution of the angle between the shower direction reconstructed using the SDM data and using the RF spectrum (3-d angle) is presented. The distribution is fitted well with a Gaussian with median $5.8^{o}$. 

\begin{figure}[h!]
\centering
\includegraphics[width=\textwidth]{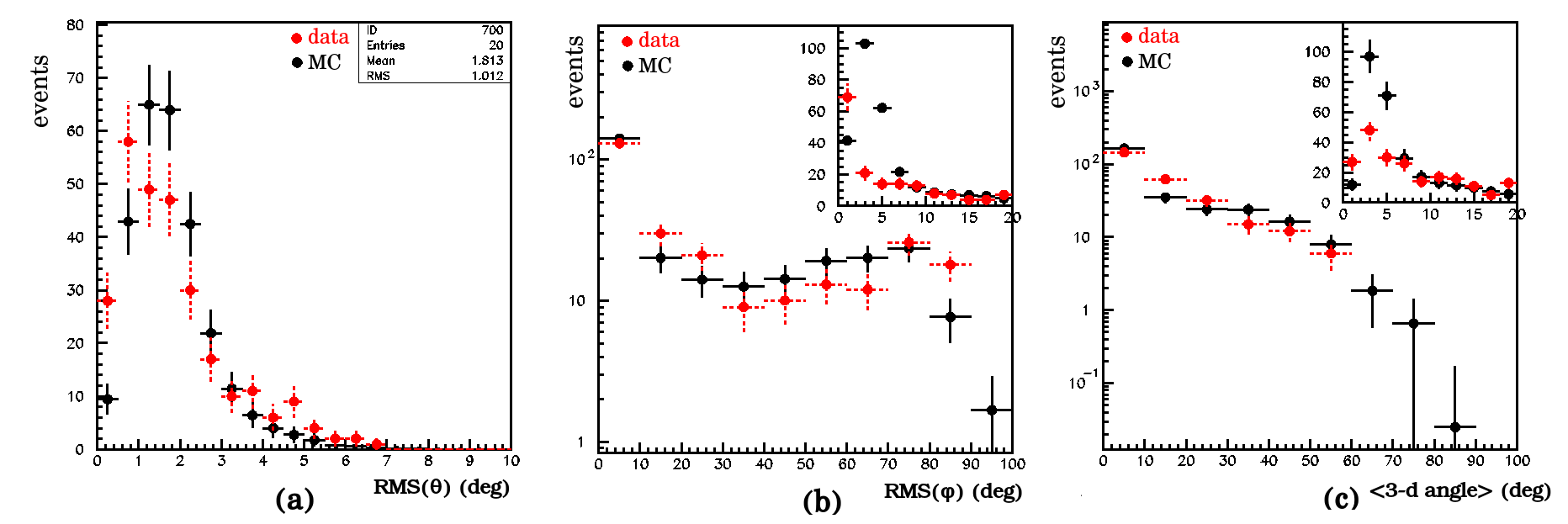} 
\caption{ Comparison between MC expectation (black points) and data (red points). {\it (a)} The distribution of the rms of the 4 antenna measurements in $\theta$. {\it (b)} The distribution of the  rms of the 4 antenna measurements in $\phi$. {\it (c)} The distribution of the mean 3d-angle between the shower directions estimated by all 4 antennas. } 
\label{fig10}
\end{figure}	 

The distributions of Figure \ref{fig9} show that the RF spectra estimated direction is highly correlated with the SDM timing estimated direction. In order to compare the RF spectra measurements independently from the particle detector data, we calculated for each event the root mean square (rms) of the 4 antenna's measurements in zenith and azimuth angle. In addition we calculated for each event the mean 3d-angle between the shower directions estimated by all 4 antennas (i.e. the mean of all 6 combinations). These distributions are shown in Figure \ref{fig10} in comparison with the MC expectation.  
The agreement between data and  MC can be considered quite satisfactory despite the deviation observed at small values. The source of this discrepancy may attributed to the poor statistics of both the data and the MC sample and/or to miss-modelling of the RF spectra.   
%  results from each antenna separately (trying to find out the efficiency of each antenna) we calculated the root mean square (rms) of the 4 antenna's measurements in zenith and azimuth angle for both data and simulations events. In figure \ref{fig10}(a) and (b) the distribution of the variance for the rms of the 4 antenna measurements in $\theta$ and $\phi$ respectively have been plotted for data (red circles) and simulations events (black). From the individual measurments of each antenna we calculate the 3-d angle between the direction as reconstructed from antennas 1-2, 1-3 and so on (6 combinations). Then the rms of the 6 combination was calculated and in figure \ref{fig10}(c) is shown the distribution of the variance for the 6 combinations rms for data (red) and mc events (black). 

%The agreement between the data and the mc simulations can be considered satisfactory despite the fact that deviate in some points. Τhe deviations noted can be attributed to the fact that in mc simulations the same shower used multiple times. This create the same mc event spectrum and does not contribute to the required statistics. Concluding we can refer that the mc simulation sample does not have the required statistics regarding the antennas spectrum.

\section{Conclusions}{\label{concl}}
\paragraph{} We have presented a method to reconstruct the shower axis of high energy showers using the spectrum of the radio signal. The reconstructed directions were compared on event by event basis with the corresponding direction calculated using the timing of the particle detectors and triangulation. For the comparison we used a sample of 274 shower events collected  by  station-A of the Astroneu array, where 4 RF antennas are triggered when a shower event is detected by the particle detectors. The comparison showed that the reconstructed directions of the 2 systems coincide within a sigma of $6.3^{o}$  and $9.4^{o}$  for the zenith and azimuth angle respectively. In addition, the simulations analysis highlighted the problems in estimating the azimuth angle with the antenna's spectrum due to the symmetries appearing in VEL. The presented results demonstrate that the RF spectrum method is efficient and applicable even in sites with strong electromagnetic noise present. 

\section*{Acknowledgments}
This research was funded by the Hellenic Open University Grant No. $\Phi$K 228: ``Development of technological applications and experimental methods in Particle and Astroparticle Physics''

%\section*{References}

\end{document}